# The Extreme Space Weather Event in 1903 October/November: An Outburst from the Quiet Sun

Hisashi Hayakawa (1-2)*, Paulo Ribeiro (3-4)**, José M. Vaquero (5-6), María Cruz Gallego (6-7), Delores J. Knipp (8-9), Florian Mekhaldi (10), Ankush Bhaskar (11-12), Denny M. Oliveira (12-13), Yuta Notsu (14-15), Víctor M. S. Carrasco (6-7), Ana Caccavari (16), Bhaskara Veenadhari (17), Shyamoli Mukherjee (17), and Yusuke Ebihara (18)

* hayakawa@kwasan.kyoto-u.ac.jp; hisashi.hayakawa@stfc.ac.uk
** pribeiro@ci.uc.pt

(1) Graduate School of Letters, Osaka University, Toyonaka, Japan.
(2) Rutherford Appleton Laboratory, Chilton, United Kingdom.
(3) CITEUC, Centre for Earth and Space Research of the University of Coimbra, Portugal
(4) Geophysical and Astronomical Observatory of the University of Coimbra, Portugal
(5) Instituto Universitario de Investigación del Agua, Cambio Climático y Sostenibilidad (IACYS), Universidad de Extremadura, Badajoz, Spain
(6) Departamento de Física, Universidad de Extremadura, Mérida, Spain
(7) Departamento de Física, Universidad de Extremadura, Badajoz, Spain.
(8) High Altitude Observatory, National Center for Atmospheric Research, Boulder, USA
(9) Smead Aerospace Engineering Sciences Department, University of Colorado Boulder, Boulder, USA
(10) Department of Geology-Quaternary Sciences, Lund University, Lund, Sweden.
(11) Catholic University of America, Washington DC, United States
(12) NASA Goddard Space Flight Center, Greenbelt, MD, United States
(13) Goddard Planetary Heliophysics Institute, University of Maryland, Baltimore County, Baltimore, MD, United States
(14) Laboratory for Atmospheric and Space Physics, University of Colorado Boulder, Boulder, USA
(15) National Solar Observatory, Boulder, CO, USA

(16) Instituto de Geofísica, Unidad Michoacan, Universidad Nacional Autonoma de Mexico, Morelia, Mexico.

(17) Indian Institute of Geomagnetism, Plot 5, Sector 18, New Panvel (West), Navi Mumbai, India

(18) Research Institute for Sustainable Humanosphere, Kyoto University, Uji, Japan

**Abstract**

While the Sun is generally more eruptive during its maximum and declining phases, observational evidence shows certain cases of powerful solar eruptions during the quiet phase of the solar activity. Occurring in the weak Solar Cycle 14 just after its minimum, the extreme space weather event in 1903 October – November was one of these cases. Here, we reconstruct the time series of geomagnetic activity based on contemporary observational records. With the mid-latitude magnetograms, the 1903 magnetic storm is thought to be caused by a fast coronal mass ejection (~1500 km/s) and is regarded as an intense event with an estimated minimum Dst' of ~−513 nT The reconstructed time series has been compared with the equatorward extension of auroral oval (~44.1° in invariant latitude) and the time series of telegraphic disturbances. This case study shows that potential threats posed by extreme space weather events exist even during weak solar cycles or near their minima.

**1. Introduction**

The Sun occasionally causes magnetic storms as a consequence of interplanetary coronal mass ejections (ICMEs) with southward interplanetary magnetic field (IMF) (*e.g.*, Gonzalez et al., 1994; Daglis et al., 1999). Due to the growing dependence on technology-based infrastructure, our civilization is increasingly vulnerable to such space weather events. Recent analyses have estimated the possible effects of extreme magnetic storms to be potentially catastrophic to the modern civilization, especially when they are as extreme as those in 1859 September and 1921 May (*e.g.*, Baker et al., 2008; Riley et al., 2018).

Statistical studies have revealed that such extreme space weather events tend to occur around the maximum and in the declining phase of solar cycles (e.g., Lefevre et al., 2016; Meng et al., 2019). Indeed, the most extreme space weather events in the observational history (Dst' ≤ −500 nT), such as those in 1859, 1872, 1909, 1921, and 1989 (*e.g.*, Allen et al., 1989; Cliver and Dietrich, 2013; WDC for Geomagnetism, Kyoto, et al., 2015; Hayakawa et al., 2018, 2019a; Love et al., 2019a, 2019b) as well as the recent Halloween sequence in 2003 (*e.g.*, Gopalswamy et al., 2005), have appeared in these phases of their corresponding solar cycles and made us wary of the Sun around the maximum to the declining phase.

However, observations show that even the quieter Sun can cause significant space weather events (*e.g.*, Kilpua et al., 2015). The extreme storm in 1986 February occurred around the solar minimum with an intensity of Dst = −307 nT (*e.g.*, Garcia and Dryer, 1987). The extreme storm of 1967 May (Dst = − 387 nT) in the ascending phase of solar cycle 20 produced significant societal impacts (Knipp et al., 2016).

Exactly a century before the Halloween sequence in 2003, another 'Halloween event' caused a significant magnetic disturbance produced geomagnetically induced currents (GICs, potentially harmful to modern power equipment and transmission lines) at mid-latitudes, resulting in the earliest documented communication networks disturbance in the Iberian Peninsula (Ribeiro et al., 2016). Interestingly, this storm occurred in the ascending phase of a weak Solar Cycle 14. Indeed, despite its lowest amplitude since the Dalton Minimum (see Clette and Lefevre, 2016), Solar Cycle 14 hosted two major space weather events in 1903 (Ribeiro et al., 2016) and 1909 (Hayakawa et al., 2019a; Love et al., 2019a).

Here we analyze the space weather events in 1903 October/November from the solar photosphere to the ground terrestrial magnetic field. We first review the solar observational data around 1903 October/November and its flare onset on the basis of contemporary solar photospheric observations and magnetic measurements. We estimate the parameters of the source ICME on the basis of the propagation time and amplitude of the storm sudden commencement (SSC). We then locate and analyze four mid-latitude magnetograms and reconstruct the equivalence of Disturbance storm time index time series (Dst'), which allows assessment of the storm intensity. We document these results with the contemporary auroral visibility and GICs, to provide a comprehensive overview of this space weather event during the early ascending phase of a weak solar cycle.

## 2. The Solar Surface in 1903 October/November

The Sun in early 1900s was relatively quiet. With onset in 1902 January, Solar Cycle 14 reached its maximal sunspot number ~ 180 in 1907 February (Figure 1, top panel). This amplitude was the lowest since the Dalton Minimum. On the surface of this quiet Sun, the sunspot group 5098 (Figure 1, bottom panel) appeared on the eastern limb of the southern hemisphere on 1903 October 25. It consisted of a relatively large composite group (493 millionths of solar hemisphere (msh); Jones, 1955), which gradually broke up in its passage across the disc, becoming a long, irregular patch that reached the central meridian on October 31. This group disappeared from the western limb on November 6 (in *Greenwich Photo-Heliographic Results 1903*, pp. 27 – 30).

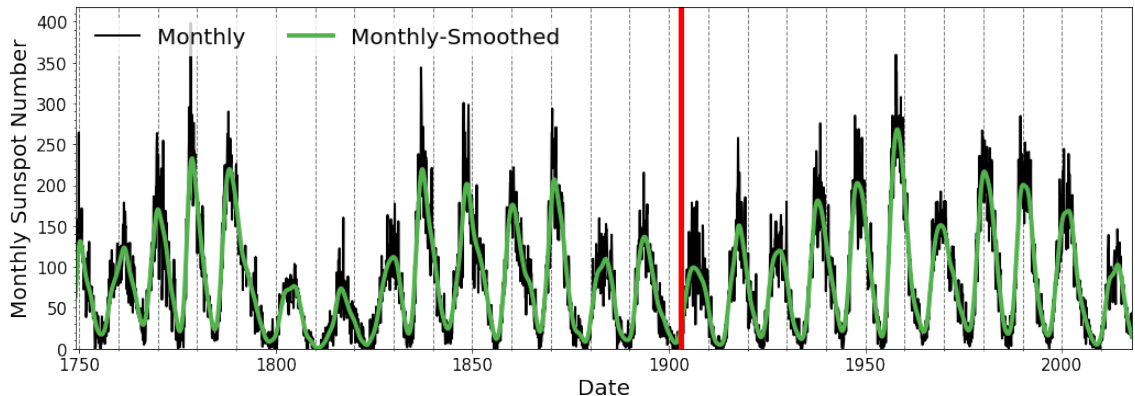

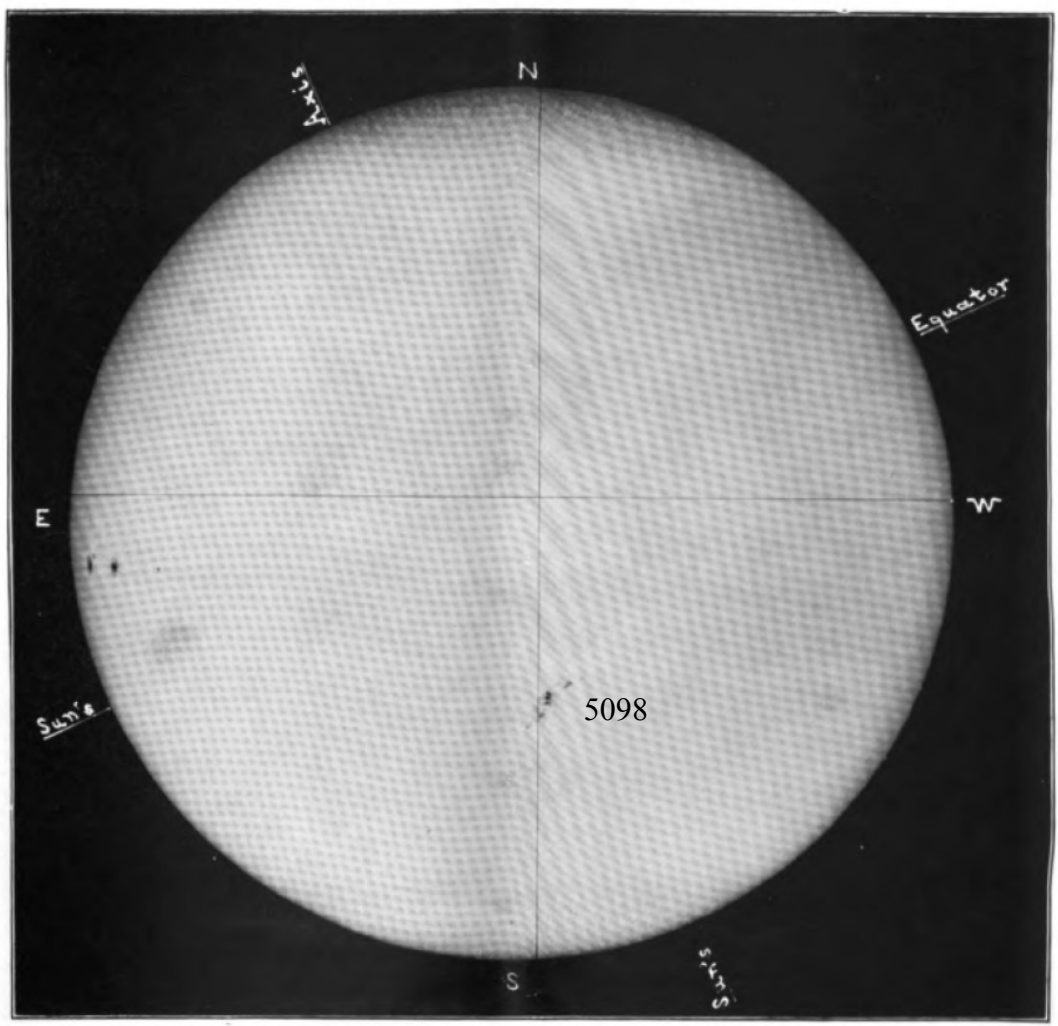

Figure 1: (Top panel) Monthly (black color) and monthly-smoothed (green color) sunspot number (version 2; see Clette and Lefevre, 2016). The vertical red line indicates the year 1903, when this study case occurred. (Bottom panel) Photograph of the Sun at 10.4 UT on 1903 October 31, taken at the Royal Observatory, Greenwich, UK, derived from Maunder (1903).

Favorably situated near the disk center, this sunspot was considerably active on 1903 October 29 – 31 (Fowler, 1903; Jones, 1955). Fowler (1903) reported "a violent distortion and reversal of the C line of hydrogen" near this group between 10 – 11 GMT on October 31. Similar reversals of the C line were seen on October 29 and 30, occasionally with more brightness but only with less distortion of the dark line, namely absorption lines (Fowler, 1903). The reversals of C line probably mean strong emission in Hα line during these flares and dynamic motion of plasma in the chromosphere (see *e.g.*, Ichimoto and Kurokawa, 1984).

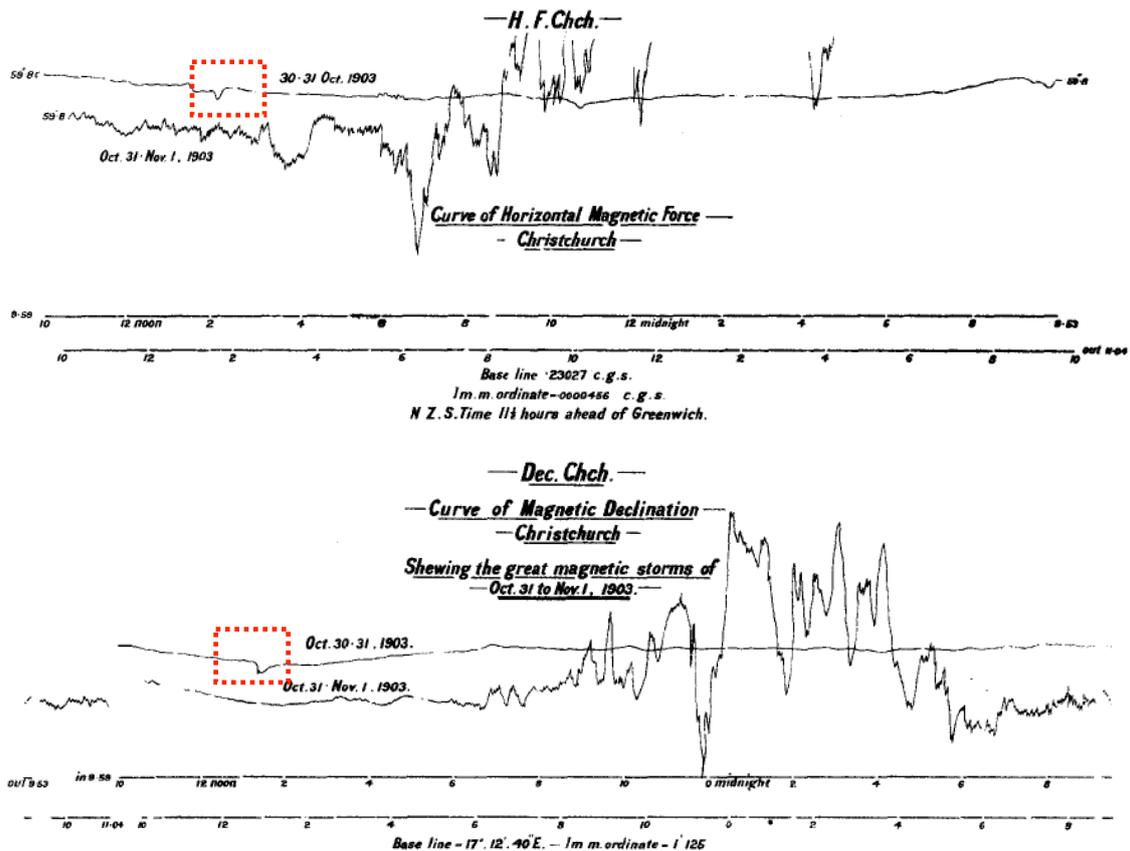

Figure 2: Magnetic crochets (enclosed by red squares) recorded in the Christchurch magnetogram (Marchant, 1904, 144/145), showing the horizontal force in the upper panel and the declination in the lower panel. The negative direction of the horizontal force is shown upward in this magnetogram.

Occurrence of intense flares during this period is confirmed with a magnetic crochet, *i.e.*, solar flare effect (*c.f.*, Jones, 1955). Figure 2 shows a magnetic crochet of ~ 15 nT at ≈ 02 GMT (13.5 local time = LT) on October 30 recorded by the Christchurch magnetogram (S43°32′, E172°37′) in New Zealand (Marchant, 1904, 144/145). The solar flare was followed by a high velocity coronal mass ejection (CME) directed to the Earth. Based on the Coimbra magnetogram in Portugal, the interplanetary CME (ICME) driving shock caused a sharp storm sudden commencement (SSC) at ≈ 5.5 GMT on October 31 with an amplitude of at least 70 nT (Ribeiro et al., 2016). However, according to the ~12 LT Colaba magnetogram in British India (based on the minute values obtained after vectorial digitizing of the analog curves), the SSC occurred at 05.35 GMT (10.85 LT) with amplitude of ≈ 98 nT (and an average increasing rate of 4.6 nT min$^{-1}$). This lets us compute the ICME propagation time as ≈ 27.5 h, slightly shorter than 28 h estimated from Zo-sé magnetogram in China (Jones, 1955), and estimate the average ICME speed as ≈ 1500 km/s. Substituting the Colaba's SSC amplitude (≈ 98 nT) into empirical equations in Araki (2014), we estimate a solar wind dynamic pressure jump of ≈ 42.7 nPa. Assuming that the solar wind consists mostly of protons, the downstream solar wind density is estimated to be ≈ 11.4 cm$^{-3}$.

Interestingly, the magnetograms at Colaba (Figure 3) and Coimbra (Figure 4 of Ribeiro et al., 2016) show sudden impulses after 20.5 GMT on 1903 October 31. These impulses suggest this storm was probably even more complex in its structure. They are presumably are due to sudden change in solar wind dynamic pressure indicating compression of magnetosphere, shock/sheath or ICME before the main ICME, as are the cases with extreme storms in 1967 and 1989 (Knipp et al., 2016; Boteler, 2019).

## 3. Magnetic Observations in 1903

After the SSC and variations of the initial phase, great magnetic disturbances were reported globally. However, many of the stations saw their recordings interrupted or incomplete due to off-scale problems associated with the fast and extreme amplitude of magnetic oscillations. The standard disturbance storm time (Dst) is a global index used to measure the geomagnetic activity and assess the intensity of magnetic storms. The index is derived from magnetograms of horizontal force (H) recorded at four middle to low latitude standard stations (Kakioka, Japan; Hermanus, South Africa; San Juan, Puerto Rico; Honolulu, Hawaii) (Sugiura, 1964). With the aim of assessing the severity of the 1903 storm, we firstly attempted to obtain the magnetograms of the historical stations closest to the ones used in the calculation of Dst'. Unfortunately, nearby surrogates for each standard station were either off scale or not in operation.

We therefore surveyed magnetic observations in four mid- to low-latitude stations with a fairly even longitudinal distribution around the Earth. We found a rather complete set of recordings and hourly data for the following observatories, for which we computed their magnetic latitude (MLAT) and longitude (MLON) in 1903 with IGRF12 model (Thebault et al., 2015), as summarized in Table 1.

| Observatory | Geographic Lat. | Geographic Long. | MLAT | MLON | Time difference | Max $\Delta H$ range | Reference |
|---|---|---|---|---|---|---|---|
| COI | N40°13′ | W8°25′ | N45.0° | E69.9° | ≈ GMT±0 | 707 | R16 |
| CLA | N18°54′ | E72°49′ | N9.9° | E143.4° | ≈ GMT+5 | 511 | IIG |
| CUA | N20°53′ | W100°53′ | N30.4° | W35.2° | ≈ GMT−7 | 570 | UNAM |
| ZKW | N31°13′ | E121°26′ | N20.0° | E170.7° | ≈ GMT+8 | 636 | Z06 |

Table 1: The reference stations used in this article: COI (Coimbra), CLA (Colaba), CUA (Cuajimalpa), and ZKW (Zi-Ka-Wei). MLAT and MLON stand for magnetic latitude and magnetic longitude, respectively. The time difference is shown referencing the Greenwich Mean Time (GMT), as defined in each observatory. The maximum range is shown in spot value with latitudinal weighting. The reference column shows where these data and details are derived from: R16 (Ribeiro et al., 2016), Z06 (Zi-Ka-Wei, 1906, pp.38-39), IIG (Indian Institute of Geomagnetism), and UNAM (Universidad Nacional Autónoma de México). The value is converted from mm to nT, according to their scale values: 7.7 nT/mm (Coimbra; Ribeiro et al., 2016), 17 nT/mm (Cuajimalpa), and 5.12 and 4.72 nT/mm (Colaba: October and November; Moos, 1910).

To obtain the hourly averages from the analog magnetograms, we traced the curves with vector-graphic programs and converted their amplitude from mm to nT. For Coimbra (COI), after the vectorization and digital reconstruction, we printed the magnetic curves (keeping the scale values) and measured the hourly mean values of H by hand (following the procedure that was commonly used for reading the classic analog magnetograms). For Colaba (CLA) and Cuajimalpa (CUA) the hourly means were obtained by simply averaging minute (CLA) and quasi-minute (CUA) values obtained during the digitization procedure. For Zi-Ka-Wei (ZKW), we have only the published tables presumably with the hourly spot values, and therefore we used these as an equivalence of hourly averages. To obtain more consistent time series with ZKW, hourly data from COI, CLA, and CUA observatories were calculated as hour-centered averages (*i.e.,* 00:00, 01:00, 02:00, *etc.*), allowing a properly averaging in the Dst' estimate.

Note that Colaba's original magnetogram shows a broken behavior, with simultaneous instrumental jumps of the H curve and its baseline (Figure 3). To reconstruct the natural trace of H we assumed the continuity of the respective baseline. In the present reconstruction of time series of Colaba, we need to carefully compare the original magnetograms and the reconstruction in Moos (1910), which shows a gap in the H-recording. In this regard, we narrowly inspected the copies of the original curves (Figure 3), and we estimated the duration of the referred gap, on the basis of the length of each baseline bar (corresponds to 2 hours of recording) and the inserted handwritten notes on the start and end times of the record. Our measurement shows that the H recording in the upper panel of Figure 3 ends at ~13.7 LT and restarts in the lower panel at ~15.2 LT, resulting in a data gap of ~1.5h hours between 8.7 GMT and 10.2 GMT.

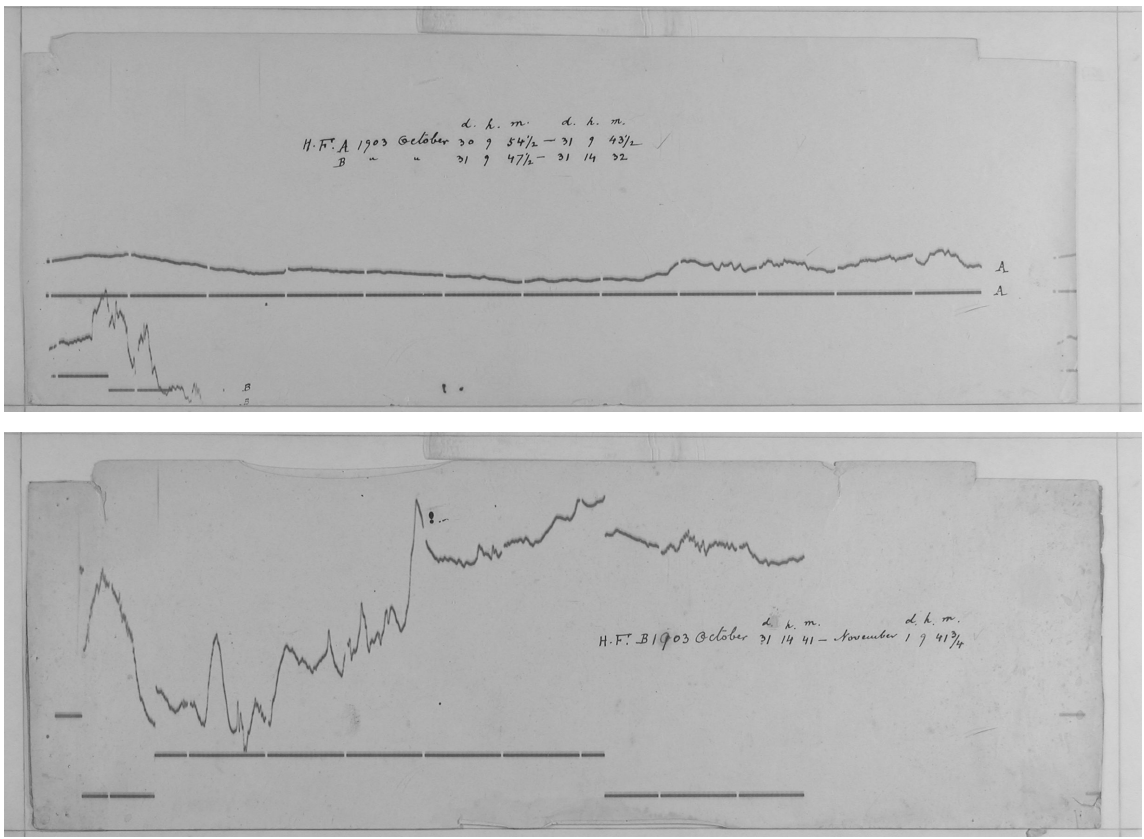

Figure 3: Original Colaba magnetogram on 1903 October 31 – November 1. Each bar of baseline shows a duration of 2 hours after the record start at 10 LT (Moos, 1910, p. 251). We assumed the continuity of baseline to reconstruct the magnetogram. (Courtesy of Indian Institute of Geomagnetism).

**4. Time Series and Intensity of the 1903 October/November Storm**

The obtained hourly averages of H for each analog magnetograms in Coimbra, Colaba, and Cuajimalpa were compared with the corresponding tabulated hourly values found for the Zi-Ka-Wei observatory. As shown in Sugiura (1964), the disturbance at each observatory is defined as:

$$D_o(t) = H_o(t) - B_o - Sq_o(t)$$

Here, the subscript 'o' refers to each observatory, and $H_o$, $B_o$, and $Sq_o$ stand for observed H, baseline of H, and solar quiet daily variation as quasi-daily variation of H, respectively. We approximated $B_o$ with the pre-storm level, $H$ hourly value at 16.5 GMT of October 30, as corresponding to the calm period before the arrival of the storm in each station. We also approximated the Sq variation of each station with the average daily variation of 5 quiet days of October 1903 (21, 20, 9, 24, 16), which were selected based on the revised daily Aa index (Lockwood et al., 2018). We then averaged weighted $D_o(t)$ of each observatory with their MLATs ($\lambda$), and obtained their average as a Dst' estimate. Figure 4 shows the hourly $D_o(t) / \cos\lambda$ of the reference stations, Coimbra, Colaba, Cuajimalpa, and Zi-ka-wei, as well as the Dst' time series as their average.

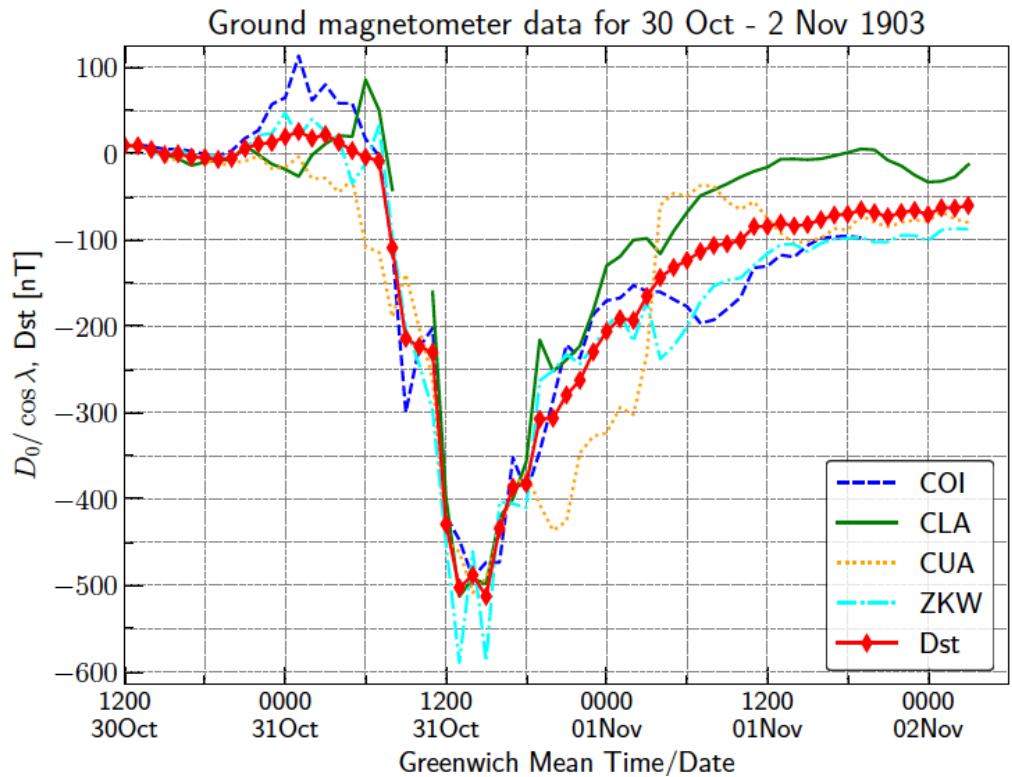

Figure 4: The plot shows $D_0$ on 1903 Oct. 30 – Nov. 2 of the reference stations, Coimbra (COI), Colaba (CLA), Cuajimalpa (CUA), and Zi-ka-wei (ZKW), as well as reconstructed Dst' estimate. As the Colaba magnetogram is scaled off at 9 – 10 h GMT, the Dst' estimate in this period is reconstructed with data from three stations. Their background data are shown in the supplementary file.

After the SSC around 5.5 GMT on October 31, the Dst' time series shows a sharp decrease from ~06 GMT, reaching its minimum −513 nT at 15 GMT. The storm main phase seems to have ended by ~ 16 GMT, and a relatively long recovery phase followed it. Contemporary estimates based on off-scaled magnetograms of Tokyo, Cheltenham (Maryland, USA), and Baldwin (Kansas, USA) point to amplitudes with latitudinal weighting of 571 nT, 805 nT, and 1010 nT, respectively (Bauer, 1904; Okada, 1904). In addition to confirming our estimate, these additional data suggest that the storm may have been even more intense.

The minimum Dst' value of −513 nT obtained for the 1903 storm ranks between the largest (1989 March; −589 nT) and the second largest (1959 July; −429 nT) magnetic storm of the official Dst index in the post-International Geophysical Year 1957 – 1958 interval. It should be highlighted that this extreme storm occurred at the onset of the weak Solar Cycle 14, while the other well known five extreme storms (Dst'/Dst ≤ −500 nT, 1859, 1872, 1909, 1921, and 1989) occurred around the maximum and in the declining phases of their corresponding solar cycle (Tsurutani et al., 2003; Cliver and Dietrich, 2013; Hayakawa et al., 2018, 2019a, 2019b; Love et al., 2019).

**5. Consequence of the Extreme Storms, Aurorae and Space Weather Hazards**

This magnetic storm caused great auroral displays and space weather hazards. The aurorae were widely seen in the territories of the Russian Empire, Australia, New Zealand, and the United States (Figure 5). The auroral visibility was reported down to Irkoutsk (Russia; N40.9° MLAT) and Walcha (Australia; S39.4° MLAT) in northern and southern hemispheres (OPCN, 1906; The Walcha Witness and Vernon County Record, 1903-11-07, p. 2). As the aurora was reported overhead at Sydney, Australia (-42.2° MLAT) (Lockyer, 1903), the footprint of the magnetic field line for its equatorward boundary of the auroral oval is conservatively reconstructed as ~ 44.1° invariant latitude (ILAT), according to the procedure in Hayakawa et al. (2018). This is almost consistent with the auroral displays in the American sector, reported overhead at Leadville, CO (*Herald Democrat*, 1903-11-01, p. 2; 47.9° MLAT) and covering all of the sky at Yerkes Observatory, WI (Barnard, 1910; 53.1° MLAT). On the other hand, the aurorae were not significantly reported in the European sector, probably because the storm main phase occurred around 6 – 16 h GMT, *i.e.*, during daytime. The European observers saw aurorae probably around the late storm recovery phase.

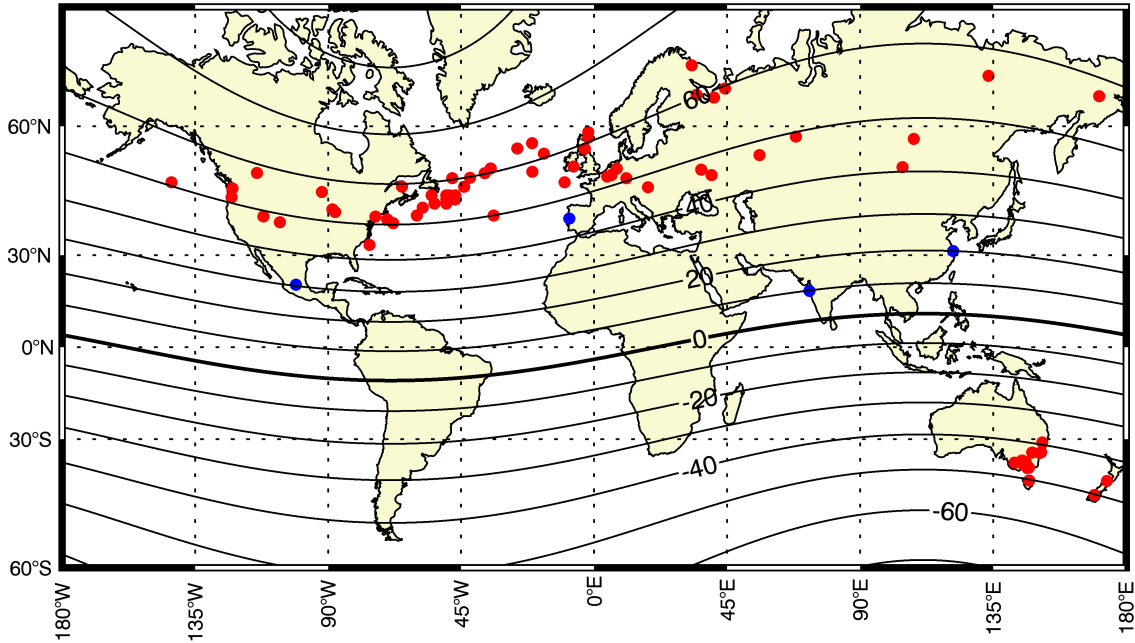

Figure 5: Auroral visibility between 1903 October 30 and November 1. The red dots show auroral observational sites in this interval of time (see Appendix 1), whereas blue dots show the reference geomagnetic stations we used in this study (see Table 1). The magnetic latitude is computed on the basis of dipole assumption of IGRF12 model (Thebault et al., 2015).

As also shown in Ribeiro et al. (2016), the telegraph communication network was interrupted in the Iberian Peninsula during 9.5 – 21 GMT, with its maximum intensity occurring during 12.5 – 15 GMT. This maximum disturbance coincides exactly with the negative peak of the Dst' time series during 12 – 16 GMT, where the Dst' intensity surpassed < −400 nT (see Figure 4). Likewise, the communications from Paris to North America and Mediterranean countries had been reportedly interrupted from ~ 9 GMT to sunset, although with a temporary recovery of normal operating conditions between ~ 16.75 – 17.5 GMT (Lockyer, 1903). This interruption mostly coincides with the period with its Dst' value more negative than −200 nT.

The GICs hit London and disturbed its railway system, telegraph connections with Latin America, France, Italy, Spain, Portugal, and Algeria (Maunder, 1903). Likewise, in the United States, this storm affected telephone lines around Chicago IL with extreme voltage of 675 volts of electricity in the wires and considered "enough to kill a man" (The Chicago Sunday Tribune, 1903-11-01, p. 8). In New South Wales of Australia, where aurora was reported overhead (Lockyer, 1903), telegraph disturbances were reported at least between 6 – 10.25 GMT (Klotz, 1904). Klotz (1904) reported "The telegraph lines running in a southerly direction were most violently affected".

## 6. Discussion and Concluding Remarks

In this letter, we aimed to provide a comprehensive view of the extreme storm of 1903 October 31, by analyzing data of the causal chain between solar photosphere to the ground terrestrial magnetic field. The Sun was rather quiet in 1903, during the second year of the ascending phase of the weak Solar Cycle 14.

Nonetheless, a relatively large composite sunspot group (5098) appeared on the eastern limb of the southern hemisphere on 1903 October 25, evolving gradually in its passage across the disc until becoming a long and irregular patch upon arrival at the central meridian on October 31. The apparent complex morphological evolution of this active region was accompanied by a set of highly energetic flares between October 29 and 31 (Fowler, 1903). In particular, the flare at ~ 02 GMT on October 30 was intense enough to be recorded as a magnetic crochet in the Christchurch magnetogram (Figure 2).

The related ICME hit the magnetosphere ~28 hours later, with the shock being recorded in the magnetograms of Coimbra and Colaba as a strong SSC around 5:30 GMT on October 31. According to our estimate, the ICME propagated into the interplanetary space with an average speed of $\approx$ 1500 km/s, and a solar wind pressure increment and density of $\approx$ 42.7 nPa, and $\approx$ 11.4 cm$^{-3}$, respectively.

In addition, the interplanetary magnetic field was strongly southward as suggested by the great storm recorded by magnetograms of four observatories at mid-MLATs (Coimbra, Cuajimalpa, Colaba, and Zi-ka-wei). On this basis, an alternative Dst' time-series has been reconstructed for the 1903 storm (Figure 4), showing that the storm's main phase lasted for almost 10 hours and reached a maximum negative value of $\approx$ −513 nT, which ranks between the largest (1989 March; −589 nT) and the second largest (1959 July; −429 nT) magnetic storms within the official Dst index.

This extreme storm caused significant auroral displays and space weather hazards. Aurorae were reported at least down to ~ 40° MLAT in both hemispheres and the equatorward boundary of auroral oval has been conservatively reconstructed at 44.1° ILAT. The telegraph and telephone lines in France, Iberian Peninsula, and even Algeria were significantly affected mostly during the storm main phase. At London, the railway system was also affected. At Chicago, an extreme voltage level of ~ 675 volts associated with extreme GICs were reported as "enough to kill a man".

It is possible this ICME was accompanied by solar energetic particles (SEPs). A preliminary survey of the Greenland ice core data from NGRIP and Dye-3 shows an enhancement in $^{36}$Cl concentrations in the early 1900s. However, the $^{10}$Be data show only a small peak using the residuals obtained by subtraction of the solar 11-year cycle in the same ice cores (Beer et al., 1990; Berggren et al., 2009; McCracken and Beer, 2015; Mekhaldi, 2019). This may indicate that the SEP associated with the CME was not large enough to produce enough $^{10}$Be as opposed to $^{36}$Cl, or the ICME did not direct a SEP event at Earth (see *e.g.*, Gopalswamy et al., 2012; Usoskin and Kovaltsov, 2012). However, this needs to be treated with caution until additional ice core data can complement these results and can rule out system effects that sometimes lead to coincidental peaks.

Although we are aware of the typical concentration of extreme space weather events around the maximum to the declining phase of solar cycles (*e.g.*, Lefevre et al., 2016), the Sun is capable of launching highly geo-effective ICMEs which in turn result in extreme space weather events even during its quiet phase, and even for a weak solar cycle, like Solar Cycle 14. Anyone who makes or uses space weather forecasts should be aware the potential of extreme space weather events even as the Sun transitions from solar minimum to the upcoming Solar Cycle 25.

**Acknowledgement**

We thank World Data Center for Geomagnetism, Kyoto for providing the observational results from Tokyo, Vieques, and Honolulu, as well as the scale unit of Colaba magnetograms, Tony Hurst and Tanja Petersen for scale unit of Christchurch magnetogram, Sunspot Index and Long-term Solar Observations (SILSO) for providing the international sunspot number, the University of Coimbra for providing Coimbra magnetograms, WDC Kyoto Geomagnetism for providing Colaba magnetograms, and Universidad Nacional Autónoma de México (especially Juan Esteban Hernández Quintero, Gerardo Cifuentes-Nava, and Armando Carrillo-Vargas) for providing Cuajimalpa magnetograms. CITEUC is funded by National Funds through FCT - Foundation for Science and Technology (project: UID/Multi/00611/2013) and FEDER -

European Regional Development Fund through COMPETE 2020 - Operational Programme Competitiveness and Internationalization (project: POCI-01-0145-FEDER-006922). Authors, BV, SM, and YE are supported by Indo-Japan research project funded by Department of Science and Technology, Govt. of India and JSPS, Japan (DST/INT/JSPS/P-137/2012). This study is one of results of the projects HISTIGUC (PTDC/FER - HFC/30666/2017) and MAG-GIC (PTDC/CTA - GEO/31744/2017), the JSPS grant-in-aids JP15H05816 (PI: S. Yoden) and JP17J06954 (PI: H. Hayakawa), and the Department of Economy and Infrastructure of the Junta of Extremadura through project IB16127 and grant GR18097 (co-financed by the European Regional Development Fund), and by the Ministerio de Economía y Competitividad of the Spanish Government (CGL2017-87917-P). AB was supported by the NASA Van Allen Probes Mission. YN was supported by JSPS Overseas Research Fellowship Program. DJK was partially supported by AFOSR grant No: FA9550-17-1-0258. Some of this material is based upon work supported by the National Center for Atmospheric Research, which is a major facility sponsored by the National Science Foundation under Cooperative Agreement No. 1852977. We gratefully thank Bruno Besser for providing Hungarian records with translations and Atsuki Shinbori for his advice on the evaluation of Sq variation. We also acknowledge the anonymous reviewer for his/her discussion and suggestions that helped to improve our manuscript.

**Appendix 1 : Historical Sources of Auroral Observations**

Monthly Weather Reports, v. 31, p. 593

Astrophysical Journal, , v. 31, pp. 209-213.

Nature, v. 69, pp. 9 and 158

Journal of the British Astronomical Association, v. 14, pp. 31-32

Annals of the Observatory of Lucien Libert, v. 11, p. 31

Popular Astronomy, v. 12, p. 288

Ciel et Terre, v. 24 , pp. 420-420

Los Angeles Times 1903-11-01        2

Astronomische Nachrichten, v. 164, pp. 77 and 355.

Időjárás, v. 7, pp. 346-349.

Uránia, v. 4, pp. 476-478.

Gippsland Times, 1903-11-02, p. 3

The Riverine Herald, 1903-11-02, p. 2

Daily Telegraph, 1903-11-02, p. 2

Traralgon Record, 1903-11-03, p. 2

The Horsham Times, 1903-11-03, p. 2

Camden News, 1903-11-05, p. 4

The Broadford Courier and Reedy Creek Times, 1903-11-06, p. 2

The North Eastern Ensign, 1903-11-06, p. 2

The Walcha Witness and Vernon County Record, 1903-11-07, p. 2

The Grenfell Record and Lachlan District Advertiser, 1903-11-07, p. 2

Cromwell Argus, 1903-11-03, p. 1

Marlborough Express, 1903-11-03, p. 1

Herald Democrat, 1903-11-01, p. 2

The Chicago Sunday Tribune, 1903-11-01, p. 8